\documentstyle[aps
,epsfig%
,preprint%
]{revtex}
\begin{document}
\draft
\preprint{Guchi-TP-009}
\date{\today
}
\title{
Conformal Quantum Mechanics in Two Black Hole Moduli Space
}
\author{Kenji~Sakamoto${}^{1}$%
\thanks{e-mail: {\tt
b1795@sty.cc.yamaguchi-u.ac.jp}
} and
Kiyoshi~Shiraishi${}^{1,2}$%
\thanks{e-mail: {\tt
shiraish@sci.yamaguchi-u.ac.jp}
}}
\address{${}^1$Faculty of Science, Yamaguchi University\\
Yoshida, Yamaguchi-shi, Yamaguchi 753-8512, Japan
}
\address{${}^2$Graduate School of Science and Engineering,
Yamaguchi University\\
Yoshida, Yamaguchi-shi, Yamaguchi 753-8512, Japan
}
\maketitle
\begin{abstract}
We discuss quantum mechanics in the moduli space consisting of 
two maximally charged dilaton black holes. The quantum mechanics 
of the two black hole system 
is similar to the one of DFF model, and this system has
the $SL(2,R)$ conformal symmetry. Also, we discuss 
the bound states in this system. 
\end{abstract}

\section{Introduction}
Recently the study of black hole moduli space has attracted much attention. 
Quantum black holes have been studied by means of quantum fields 
or strings interacting with a single black hole. 
In the past few years the new quantum mechanics of an arbitrary number $n$ of 
supersymmetric black holes has been focused. 
Configurations of $n$ static black holes parametrize a moduli space.
The low-lying quantum states of the system are governed by quantum mechanics 
in the moduli space.
The effective theories of quantum mechanics in moduli space of
Reissner-Nordstr\"{o}m multi-black holes were constructed 
in \cite{Gibbons}. 
Recently, (super)conformal quantum mechanics are 
constructed in moduli space of 
four- and five-dimensional 
multi-black holes in the near-horizon limit in \cite{jmas}, \cite{as2}.

The motivation behind these works is the hope that information 
of quantum states will lead to the black hole entropy. 
The quantum states supported in the near-horizon region 
can be interpreted as internal states of black holes, 
and the number of such states is related to the black hole entropy.
The other motivation is the hope that investigating these moduli spaces will 
lead to the understanding of $\rm{AdS}_2/\rm{CFT}_1$ correspondence 
and unravel some novel features in multi-black hole mechanics.

The geometry of black hole moduli spaces was first discussed by Ferrell and 
Eardley in four dimensions \cite{fe}. 
Further the black hole moduli spaces geometry with dilaton coupling 
in $N+1$ dimensions was discussed by one of the present authors \cite{shir}. 
In these works the structure of the moduli space geometry is found to be 
different for dimensions and values of dilaton coupling. 

In this paper, we discuss quantum mechanics in the moduli space of 
two maximally charged dilaton black holes. 
In section 2, we discuss the moduli space structure of two black hole system 
with dilaton coupling in any dimensions. 
In section 3, we find that the quantum mechanics on this moduli space has 
conformal symmetry in close limit, where multi-black holes approach to each other. 
We give explicit representation of the dilatational generator $D$ and conformal 
generator $K$. 
Then we study the effect on conformal 
quantum mechanics from the different structures of the moduli space geometry. 
In section 4, we consider the bound states by utilizing the moduli space structure 
for the particular dilaton coupling value in fixed dimensions. 
In section 5, we will give discussion and conclusion.

\section{The Moduli Space Metric for the System of Maximally Charged 
Dilaton Black Holes}
The Einstein-Maxwell-dilaton system contains a dilaton field $\phi$ coupled to a $U(1)$ gauge field $A_{\mu}$ beside the Einstein-Hilbert gravity.
In the $N+1$ dimensions $(N\ge3)$, the action for the fields with particle sources is 
\begin{eqnarray}
S&=&\int d^{N+1}x \frac{\sqrt{-g}}{16\pi} \left[ R -
\frac{4}{N-1} (\nabla \phi)^{2} - e^{-\{4a/(N-1)\}\phi} F^{2} \right] \nonumber \\
& & \hspace{0.5cm} -\sum^n_{i=1} \int ds_i \left( m_ie^{-\{4a/(N-1)\}\phi} +Q_iA_{\mu}\frac{dx_i^{\mu}}{ds_i}\right),\label{eq:action}
\end{eqnarray}
where $R$ is the scalar curvature and $F_{\mu\nu}=\partial_\mu A_\nu-\partial_\nu A_\mu$. We set the Newton constant $G=1$. 
The dilaton coupling constant $a$ can be assumed to be a positive value.

The metric for the $n$-body system of maximally-charged dilaton black holes has been known as \cite{shir}
\begin{equation}
ds^2=-U^{-2}({\bf{x}}) dt^2+U^{2/(N-2)}({\bf{x}})d{\bf{x}}^2, 
\end{equation}
where
\begin{eqnarray}
U({\bf{x}})&=&(F({\bf{x}}))^{(N-2)/(N-2+a^2)}, \label{eq:U} \\
F({\bf{x}})&=&1+\sum^n_{i=1}\frac{\mu_i}{(N-2)|{\bf{x}}-{\bf{x}}_i|^{N-2}}. \label{eq:F}
\end{eqnarray}

Using these expressions, the vector one form and dilaton configuration are written as
\begin{eqnarray}
A=\sqrt{\frac{N-1}{2(N-2+a^2)}}\left(1-\frac{1}{F({\bf{x}})}\right)dt, \\
e^{-4a\phi/(N-1)}=(F({\bf{x}}))^{2a^2/(N-2+a^2)}.
\end{eqnarray}
In this solution, the asymptotic value of $\phi$ is fixed to be zero.

The electric charge $Q_i$ of each black hole are associated with the corresponding mass $m_i$ by
\begin{eqnarray}
m_i&=&\frac{A_{N-1}(N-1)}{8\pi(N-2+a^2)}\mu_i,\\
|Q_i|&=&\sqrt{\frac{N-1}{2(N-2+a^2)}}\mu_i, 
\end{eqnarray}
where $A_{N-1}=2\pi^{N/2}/\Gamma(\frac{1}{2}N)$.

We consider that the perturbed metric and potential can be written in the form 
\begin{eqnarray}
ds^2=-U^{-2}({\bf{x}})dt^2+2{\bf{N}}d{\bf{x}}dt+U^{2/(N-2)}({\bf{x}})d{\bf{x}}^2, \\
A=\sqrt{\frac{N-1}{2(N-2+a^2)}}(1-\frac{1}{F({\bf{x}})})dt+{\bf{A}}d{\bf{x}}, 
\end{eqnarray}
where $U({\bf{x}})$ and $F({\bf{x}})$ are defined by (\ref{eq:U}) and (\ref{eq:F}). 
We have only to solve linearized equations with perturbed sources up to $O(v)$ 
for $N_i$ and $A_i$. 
(Here $v$ represents the velocity of the black hole as a point source.) 
We should note that each source plays the role of a maximally charged dilaton black hole.

Solving the Einstein-Maxwell equations and substituting the solutions, 
the perturbed dilaton field and sources to the action (\ref{eq:action}) with proper boundary terms, 
we get the effective Lagrangian up to $O(v^2)$ for $n$-maximally charged dilaton black hole system
\begin{eqnarray}
L&=&-\sum_{i=1}^n m_i +\sum_{i=1}^n\frac{1}{2}m_i({\bf{v}}_i)^2 \nonumber \\
& & \vspace{1cm} + \frac{(N-1)(N-a^2)}{16\pi(N-2+a^2)^2} 
\int d^N x (F({\bf{x}}))^{2(1-a^2)/(N-2+a^2)}\sum_{i,j}^n
\frac{({\bf{n}}_i\cdot{\bf{n}}_j)|{\bf{v}}_i-{\bf{v}}_j|^2\mu_i\mu_j}
{2|{\bf{r}}_i|^{N-1}|{\bf{r}}_j|^{N-1}} , \label{lag1}
\end{eqnarray}
where ${\bf{r}}_i={\bf{x}}-{\bf{x}}_i$ and ${\bf{n}}_i={\bf{r}}_i/|{\bf{r}}_i|$. $F({\bf{x}})$ is defined by (\ref{eq:F}).
In general, a naive integration in equation (\ref{lag1}) diverges. Therefore, we 
regularize that divergent terms proportional 
to $\int d^Nx\delta^n(x)/|x|^p \ (p>0)$ 
which appear when the integrand is expanded must be regularized \cite{InPl}. We set them to zero. The prescription is equivalent to carrying out the following replacement in equation (\ref{lag1})
\begin{eqnarray}
(F(x))^{\frac{2(1-a^2)}{(N-2+a^2)}} \to -1+\left[ 1+
\frac{8\pi(N-2+a^2)}{A_{N-1}(N-2)(N-1)}
\frac{m_a}{|{\bf{r}}_a|^{N-2}} \right]^\frac{2(1-a^2)}{N-2+a^2} \nonumber \\
+\left[1+\frac{8\pi(N-2+a^2)}{A_{N-1}(N-2)(N-1)}
\frac{m_b}{|{\bf{r}}_b|^{N-2}} \right]^\frac{2(1-a^2)}{N-2+a^2}.
\end{eqnarray}
After regularization, the effective Lagrangian for two body system (consisting of black hole labeled with $a$ and $b$) can be rewritten as
\begin{eqnarray}
L_{2B}=&-&M +\frac{1}{2}M{\bf{V}}^2 \nonumber \\
&+&\frac{1}{2} \mu {\bf{v}}^2 
\Biggl[ 1-\frac{M}{\mu}
-\frac{8\pi(N-a^2)}{A_{N-1}(N-2)(N-1)}\frac{M}{r^{N-2}} \nonumber \\
&+&\frac{M}{m_a} \left( 1+\frac{8\pi(N-2+a^2)}{A_{N-1}(N-2)(N-1)}
\frac{m_a}{r^{N-2}} \right) ^{(N-a^2)/(N-2+a^2)} \nonumber \\
&+&\frac{M}{m_b} \left( 1+\frac{8\pi(N-2+a^2)}{A_{N-1}(N-2)(N-1)}
\frac{m_b}{r^{N-2}} \right) ^{(N-a^2)/(N-2+a^2)}
\Biggl], 
\end{eqnarray}
where $M=m_a+m_b$, $\mu=m_am_b/M$, ${\bf{V}}=(m_a{\bf{v}}_a+m_b{\bf{v}}_b)/M$, ${\bf{v}}={\bf{v}}_a-{\bf{v}}_b$ and 
$r=|{\bf{x}}_a-{\bf{x}}_b|$.
Since the motion of the center-of-mass is separable,
we assume that the velocity of center of mass $V$ vanishes. 
Thus the metric of the $N+1$ dimensional moduli space for two-body system is 
\begin{equation}
g_{ab}=\gamma(r)\delta_{ab},\label{mmet1}
\end{equation}
with 
\begin{eqnarray}
\gamma(r)=1&-&\frac{M}{\mu}
-\frac{8\pi(N-a^2)}{A_{N-1}(N-2)(N-1)}\frac{M}{r^{N-2}} \nonumber \\
&+&\frac{M}{m_a}\left( 1+\frac{8\pi(N-2+a^2)}{A_{N-1}(N-2)(N-1)}
\frac{m_a}{r^{N-2}}\right)^{(N-a^2)/(N-2+a^2)} \nonumber \\
&+&\frac{M}{m_b}\left(1+\frac{8\pi(N-2+a^2)}{A_{N-1}(N-2)(N-1)}
\frac{m_b}{r^{N-2}}\right)^{(N-a^2)/(N-2+a^2)}.\label{mmet2}
\end{eqnarray}

\section{Quantum mechanics in two-black hole moduli space}
We consider quantum mechanics in moduli space in this section. 
The quantization of moduli parameters has been discussed in \cite{TrFe}.
In this section, we shall reveal that quantum mechanics in moduli space has 
an $SL(2,R)$ conformal symmetry in close limit. 

Let us introduce a wave function $\Phi$ in the moduli space, which obeys the 
Schr\"{o}dinger equation
\begin{eqnarray}
i\hbar\frac{d\Phi}{dt}&=&\left(-\frac{\hbar^2}{2\mu}\nabla^2+\hbar^2\xi R_{(MS)}\right)\Phi \nonumber \\
&=&H\Phi, \label{scheq}
\end{eqnarray}
where $\nabla^2$ is the covariant Laplacian constructed from moduli space metric 
and $R_{(MS)}$ is the scalar curvature of the moduli space. 
We assume $\xi=0$ in this paper though this term may be present in most 
general case.
From the Schr\"{o}dinger equation (\ref{scheq}), we can get the operator formalism of the Hamiltonian. 

In the (3+1) dimensions case, Michelson and Strominger discussed the moduli space 
quantum mechanics \cite{jmas}. 
The multi-black hole moduli space has two types of noncompact regions, 
asymptotically flat region and near horizon region. 
In these regions, the wave function $\Phi$ is defined. 
Noncompact regions prevent the existence of a normalizable ground state. 
The asymptotically flat region correspond to widely spreaded black holes. 
And the near horizon region correspond to near-coincident black holes. 
In a low-energy and close limit, this 
asymptotically flat region decouple from the near-horizon region. 
Furthermore, the noncompact region for near horizon region can be eliminated by 
a generalization of the DFF suggestion \cite{dff}. 
The result in close limit is that a normalizable ground state 
and well defined quantum mechanics can be considered. 
Consequently, when we consider quantum mechanics in close limit, we can 
simply find the normalizable ground state. 

In another dimension and for the other dilaton coupling, 
it seems unclear whether the conformal structure exist or not. 
If the the quantum mechanics in the moduli space has a conformal symmetry, 
the DFF suggestion is available and the well defined quantum mechanics can be 
constructed. 
We show that the moduli space has an $SL(2,R)$ conformal symmetry 
in close limit $r\to0$. 

In close limit $r\to0$, the moduli space metric (\ref{mmet1}) 
is shown the different behavior for each values of dilaton coupling. \\
For $a^2<1$,
\begin{eqnarray}
\gamma(r)&\approx& k_1r^{-\frac{(N-a^2)(N-2)}{N-2+a^2}},\\
k_1&\equiv& M\left( \frac{8\pi(N-2+a^2)}{A_{N-1}(N-2)(N-1)}\right)^{\frac{N-a^2}{N-2+a^2}}\left(m_a^{\frac{2-2a^2}{N-2+a^2}}+m_b^{\frac{2-2a^2}{N-2+a^2}}\right). \label{k1}
\end{eqnarray}
For $a^2=1$, 
\begin{eqnarray}
\gamma(r)&\approx&k_2\frac{1}{r^{N-2}}, \\
k_2 & \equiv& \frac{8\pi M}{A_{N-1}(N-2)}.\label{k2}
\end{eqnarray}
For $a^2>1$ (here $a^2\neq N$), 
\begin{eqnarray}
\gamma(r)&\approx&k_3\frac{1}{r^{N-2}},\\
k_3&\equiv&-\frac{8\pi M(N-a^2)}{A_{N-1}(N-2)(N-1)}.\label{k3}
\end{eqnarray}
Furthermore another special case exists: 
For $a^2=N$, the moduli space metric is independent of $r$,
\begin{eqnarray}
\gamma(r)=1\equiv k_4. \label{k4}
\end{eqnarray}

From the moduli space metric in close limit, 
we can obtain the operator formalism of Hamiltonians by (\ref{scheq}). 
The Hamiltonians in the moduli space are as follows. \\
For $a^2<1$ (here except for the case of $a^2=1/3$ for $N=3$),
\begin{eqnarray}
H=-\frac{1}{2\mu k_1}[\frac{d^2}{du^2}-\frac{C_1}{u^2}], \label{H1}
\end{eqnarray}
where 
\begin{eqnarray}
C_1&\equiv& \frac{1}{4} (N-3)(N-1)+\left(\frac{2(N-2+a^2)}{N^2-(4+a^2)N+4}\right)^2l(l+1), \label{C1} \\
u&=&-\frac{2(N-2+a^2)}{N^2-(4+a^2)N+4}r^{-\frac{N^2-(4+a^2)N+4}{2(N-2+a^2)}}.\label{u1}
\end{eqnarray}
For $a^2=1$ (here $N\neq4$), 
\begin{eqnarray}
H&=&-\frac{1}{2\mu k_2}[\frac{d^2}{du^2}-\frac{C_2}{u^2}], \label{H2}
\end{eqnarray}
where
\begin{eqnarray}
C_2&\equiv&\frac{1}{4}(N-3)(N-1)+\frac{4l(l+1)}{(N-4)^2}, \label{C2} \\
u&=&-\frac{2}{N-4}r^{-\frac{N-4}{2}}. \label{u2}
\end{eqnarray}
For $a^2>1$ (here $N\neq4$), 
\begin{eqnarray}
H&=&-\frac{1}{2\mu k_3}[\frac{d^2}{du^2}-\frac{C_3}{u^2}], \label{H3}
\end{eqnarray}
where 
\begin{eqnarray}
C_3&\equiv&\frac{1}{4}(N-3)+\frac{4l(l+1)}{(N-4)^2}, \label{C3} \\
u&=&-\frac{2}{N-4}r^{-\frac{N-4}{2}}.\label{u3}
\end{eqnarray}
Other special cases exist: 
For $a^2=N$,
\begin{equation}
H=-\frac{1}{2\mu k_4}[\frac{d^2}{du^2}-\frac{C_4}{u^2}], \label{H4}
\end{equation}
where 
\begin{eqnarray}
C_4&\equiv&\frac{1}{4}(N-3)(N-1)+l(l+1), \\
u&=&r. \label{u4}
\end{eqnarray}
For $a^2>1$ and $N=4$,
\begin{eqnarray}
H&=&-\frac{1}{2\mu k_3}[\frac{d^2}{du^2}-l(l+1)], 
\end{eqnarray}
where 
\begin{eqnarray}
u&=&\log r.
\end{eqnarray}
For $a^2=(N-2)^2/N$, we have two cases, $N=3$ (then $a^2=1/3$) and $N=4$ ($a^2=1$). For $N>4$ we find $a^2>1$, then the Hamiltonian is the same as in the case of $a^2>1$.\\
For $N=3$ and $a^2=1/3$,
\begin{equation}
H=-\frac{1}{2\mu k_1}[\frac{d^2}{du^2}-l(l+1)],
\end{equation}
where
\begin{equation}
u=\log r.
\end{equation}
For $N=4$ and $a^2=1$,
\begin{equation}
H=-\frac{1}{2\mu k_2}[\frac{d^2}{du^2}-l(l+1)], 
\end{equation}
where
\begin{equation}
u=\log r.
\end{equation}

Consequently, except for the cases of $a^2=1/3$ for $N=3$ and 
$a^2\geq1$ for $N=4$, Hamiltonian (\ref{H1}), (\ref{H2}), (\ref{H3})
and (\ref{H4}) are following form: 
\begin{equation}
H=-\frac{1}{2\mu k_i}[\frac{d^2}{du^2}-\frac{C_i}{u^2}], \label{hamil}
\end{equation}
where the index $i$ runs from $1$ to $4$. 
The variables $u$ are eq.(\ref{u1}), (\ref{u2}), (\ref{u3}) and  (\ref{u4}), where 
we note that $u$ are not necessarily positive power of $r$. 
The form of this Hamiltonian is well known as DFF Hamiltonian \cite{dff}. 
The quantum mechanical system of this Hamiltonian has the conformal symmetry. 
So, in this model, the quantum mechanics in the black hole moduli space has conformal symmetry. Then the dilatational generator $D$ and special conformal generator $K$ are defined by 
\begin{eqnarray}
D&=&-\frac{i}{2}[u\frac{d}{du}+\frac{d}{du}u], \\
K&=&\frac{1}{2}\mu k_i u^2. 
\end{eqnarray}
These generators have the $SL(2,R)$ algebra, 
\begin{eqnarray}
\left[ D,H \right]&=& 2iH, \\ 
\left[ D,K \right]&=&-2iK, \\ 
\left[ H,K \right]&=&-iD.
\end{eqnarray}
Therefore the quantum mechanics in the black hole moduli space has the conformal symmetry in close limit. We consider the linear combinations 
\begin{eqnarray}
L_{\pm1}=\frac{1}{2}(\alpha H-\frac{K}{\alpha} \mp iD), \\ 
L_{0}=\frac{1}{2}(\alpha H+\frac{K}{\alpha}),
\end{eqnarray}
where $\alpha$ is a parameter with dimensions of length squared. 
In these bases the generators obey the standard $SL(2,R)$ commutation relation, 
\begin{eqnarray}
[L_{+1}, L_{-1}]=2L_{0}, \\ 
\left[ L_{0}, L_{\pm 1} \right]= \mp L_{ \pm 1}, 
\end{eqnarray}
where we choose the units such that $\alpha=1$. 
This conformal structure corresponds to the result of 
Birmingham, Gupta and Sen \cite{GS2}, 
where the Hamiltonian which has the form of eq.(\ref{hamil}) 
belongs to the enveloping algebra of the full Virasoro algebra. 

In this section, the quantum mechanics are considered 
in the case of arbitrary dimensions and values of dilaton coupling. 
Except for the special cases, these quantum mechanics have 
the same conformal structure because these Hamiltonian are in the same form. 
Although the structure of moduli space geometry is different 
for dimensions and values of dilaton coupling \cite{shir}, 
these difference of moduli space does not affect the conformal structures 
of quantum mechanics. 

If we set the dilaton coupling to zero and $N=4$ in our 
analysis based on the moduli metric, 
we reproduce the same result as dilatational 
generator and special conformal generator in the work of 
Britto-Pacumio, Strominger and Volovich \cite{as3}, 
where the quantum mechanics of slowly-moving multi-black holes 
in 4+1 dimensions is considered and 
the conformal structure is discussed in two black hole system. 
Also, if we set the dilaton coupling to zero and one black hole mass to infinity 
in the case of $N=4$ in our model, 
we reproduce the same result as $H$, $D$ and $K$ in the work 
of Claus {\it et al}. \cite{TownProe}. 
They discuss the dynamics of a particle near horizon of an extreme 
Reissner-Nordstr\"{o}m black hole. 
The mechanics of their model has the (super)conformal symmetry. 
Although our model consider the two black hole system, 
our model is reduced to the same system 
when one black hole mass is set to infinity.

\section{The Bound States in the moduli space}
Let consider in the case of Hamiltonian (\ref{hamil}). 
Although the Hamiltonian (\ref{hamil}) is defined in the moduli space, 
we can regard $H$ as the operator on Hilbert space. 
Moreover the operator $H$ belongs to a general class of objects known as 
unbounded linear differential operators on Hilbert space. 
Before considering the bound states, we summarize some properties of 
this operator along the work of Govindarajan, Suneeta and Vaidya \cite{GSV}. 

The Hamiltonian $H$ is written as
\begin{equation}
H=-\frac{1}{2\mu k}\left[ \frac{d^2}{du^2}-\frac{C}{u^2}\right],
\end{equation}
where an index $i$ of $C_i$ and $k_i$ is omitted. 
This is an unbounded differential operator defined in $R^+ \equiv[0,\infty]$.
$H$ is a symmetric operator on the domain 
$D(H)\equiv \{  \phi \in L^2[R^+,du],\ \phi(0)=\phi'(0)=0 \}$. 
It is known that for $C\geq 3/4$, $H$ is 
(essentially) self-adjoint on the domain $D(H)$. 
For $-1/4\leq C <3/4$, $H$ is not self-adjoint on the domain $D(H)$ but 
admits self-adjoint extensions, where 
the self-adjoint extensions are labeled by a $U(1)$ parameter $e^{iz}$. 
Each value of parameter $z$ defines the domain $D_{z}(H)$. 
The operator $H$ is self-adjoint in the domain $D_{z}(H)$ which contains 
all the vector in $D(H)$ and vectors of the 
form $\phi_{+}(u)+e^{iz} \phi_{-}(u)$, where
\begin{eqnarray}
\phi_{+}=(\sqrt{2\mu k}u)^{1/2}H_{\nu}^{(1)}(\sqrt{2\mu k}u e^{i\pi /4}), \\
\phi_{-}=(\sqrt{2\mu k}u)^{1/2}H_{\nu}^{(2)}(\sqrt{2\mu k}u e^{-i\pi /4}),
\end{eqnarray}
where $\nu=\sqrt{1/4+C}$, and $H_{\nu}^{(1,2)}$ are Hankel functions. 

We can now solve the eigenvalue equation for bound states, 
\begin{equation}
-\frac{d^2\Phi}{du^2}+\frac{C}{u^2}\Phi=-2\mu k E \Phi,
\end{equation}
where an index $i$ of $C_i$ and $k_i$ is omitted. 
For $C\geq3/4$, there is no bound state. 
Namely there is no normalizable solution to the 
Schr\"{o}dinger equation with negative energy. 
For $-1/4\leq C <3/4$, 
there is exactly eigenfunction
\begin{equation}
\Phi=B(\sqrt{2\mu k E_b} u)^{1/2}[J_{\nu}(i \sqrt{2\mu k E_b} u)-e^{i\pi\nu}J_{-\nu}(i \sqrt{2\mu k E_b} u)], \label{boundstate}
\end{equation}
where $J_{\nu}$ are Bessel function and $B$ is the normalization constant. 
From eq.(\ref{C1}), (\ref{C2}) and (\ref{C3}) and the condition of $-1/4\leq C <3/4$, the spatial dimension $N$ is restricted. 
In the case of s wave, the condition of $a^2$ and $N$ is as follows: 
\begin{eqnarray*}
For\ \  a^2 \leq 1&:& N=3. \\
For\ \  a^2>1&:& N =3,5.
\end{eqnarray*}
Then the bound states (\ref{boundstate}) are meaningful. 
For simplification, we consider the case of s wave in the following discussion. 

Note that the quantum system under consideration is discussed 
in close limit $r\to0$. 
So this eigenfunction is meaningful in the limit of $u$ as 
close limit $r\to0$. 
However close limit of $u$ is 
different by the condition of $a^2$ and $N$. 
For example, for $N=3$ and $a^2=1$, 
$u=2\sqrt{r}$. Then $u\to0$ corresponds to close limit of $r\to0$. 
For another example, for $N=5$ and $a^2>1$, $u=-2/\sqrt{r}$.
Then $u\to \infty$ meens close limit of $r\to 0$. 
Consequently, the close limit is separated the following cases,
\begin{eqnarray*}
For\ \  u\to 0\ &:& a^2>1/3\ for\ N=3. \\
For\ \  u\to \infty&:& a^2\leq1/3 \ for \ N=3 \ \ and\ \ a^2>1\ for\ N=5.
\end{eqnarray*}
The value of $a^2=1/3$ for $N=3$ is the critical point of moduli space 
structures \cite{shir}. Therefore we expect that the difference of moduli space 
structures affects bound states. 

The bound states (\ref{boundstate}) have the following behavior. 
In the limit of $u\to0$ 
\begin{equation}
\Phi\sim B \left[\frac{(\sqrt{2\mu k}u)^{\nu+1/2}}{2^{\nu}}
\frac{(\sqrt{E_b})^{\nu+\frac{1}{2}}e^{i\frac{\nu\pi}{2}}}{\Gamma(1+\nu)}
-\frac{(\sqrt{2\mu k}u)^{-\nu+1/2}}{2^{-\nu}}
\frac{(\sqrt{E_b})^{-\nu+\frac{1}{2}}e^{i\frac{\nu\pi}{2}}}{\Gamma(1-\nu)}
\right]. \label{limitphi}
\end{equation}
In the limit of $u\to \infty$ 
\begin{equation}
\Phi\sim B \sqrt{\frac{2}{\pi}}\sum^{\infty}_{n=0} 
\frac{(\nu,n)}{(2\sqrt{2\mu k}u)^n} 
\frac{e^{-\sqrt{2\mu k E_b}u}}{E_b^{n/2}} \sin{\nu \pi},\label{limitphiinfty}
\end{equation}
where
\begin{eqnarray}
(\nu,n)&=&\frac{(4\nu^2-1^2)(4\nu^2-3^2)\cdots(4\nu^2-(2n-1)^2)}{n!(2)^{2n}}, \\
(\nu,0)&=&1.
\end{eqnarray}

From this behavior of bound states, 
we can obtain the corresponding bound energy $E_b$. 
Because $H$ is self-adjoint on the domain $D_z(H)$ for $-1/4\leq C <3/4$, 
the eigenfunction must belong to the domain $D_z(H)$. \\
When $a^2>1/3$ for $N=3$, $u\to 0$ means close limit. 
Then 
\begin{eqnarray}
\phi_{+} (u)&+&e^{iz}\phi_{-} (u)\sim \frac{i}{\sin{\nu \pi}}
\biggl[\frac{(\sqrt{2\mu k}u)^{\nu+1/2}}{2^{\nu}}
\frac{e^{-i3\pi\nu/4}-e^{i(3\pi\nu/4+z)}}{\Gamma(1+\nu)} \nonumber \\
& & \hspace{4cm} +\frac{(\sqrt{2\mu k}u)^{-\nu+1/2}}{2^{-\nu}}
\frac{e^{i(\pi\nu/4+z)}-e^{-i\pi\nu/4}}{\Gamma(1-\nu)}
\biggr].\label{limitphi2}
\end{eqnarray}
Since the coefficients of $u$ in (\ref{limitphi}) and (\ref{limitphi2}) must be 
coincident, the bound energy $E_b$ at $a^2>1/3$ for $N=3$ is, 
\begin{equation}
E_b=\left[ \frac{\sin(z/2+3\pi \nu/4)}{\sin(z/2+\pi \nu/4)} \right]^{1/\nu}. 
\end{equation}
In the case of $a^2\leq1/3$ for $N=3$ 
and $a^2>1$ for $N=5$, $u\to \infty$ are close limit.
\begin{equation}
\phi_{+} (u)+e^{iz}\phi_{-} (u)\sim 2\sqrt{\frac{2}{\pi}} e^{iz/2} 
e^{-\sqrt{\mu k} u} 
\cos{\left[ \sqrt{\mu k} u-\frac{\nu \pi}{2}
-\frac{3\pi}{8}-\frac{z}{2}
\right]}.
\end{equation}
From this equation and (\ref{limitphiinfty}), the bound energy $E_b$ is written
\begin{equation}
E_b=\frac{1}{2},
\end{equation}
where we consider that the dumping factor dominates the behavior of function more than the fluctuation factor at $u\to \infty$. 

In this section, we have studied the bound states in moduli space. 
In the work of Govindarajan, Suneeta and Vaidya \cite{GSV} 
and Gupta and Sen \cite{GS}\cite{GS3}, 
the black hole quantum states are given in the space-time made by one black hole. 
But we now consider the two-black hole system in this paper and 
the quantum states are given in the moduli space. 
When the quantum states are considered in the moduli space, 
we can exactly treat all interaction between the multi-black holes in principle 
because the moduli space metric is constructed 
from the effective Lagrangian contained all interaction. 
So the quantum mechanics in moduli space is very meaningful. 
In this paper, however, we have considered the two black hole system. 
Therefore the interaction between two black holes is very simple and 
the difference from the other works does not appear explicitly. 

\section{Conclusion}
In this paper, we studied quantum mechanics in the moduli space consisting of two 
maximally charged dilaton black holes. 
From the effective action for the two-body system 
of maximally charged dilaton black hole in any dimensions, 
we can obtain the moduli space metric of two black hole system. 
To investigate the Hamiltonian produced from these moduli space metric, 
we discussed the quantum mechanics in these moduli spaces. 
Except for the special cases, 
these quantum mechanics are similar to the DFF model of conformal 
quantum mechanics \cite{dff}, 
and these system has the $SL(2,R)$ conformal symmetry. 
Although quantum mechanics are considered in the case of arbitrary dimensions 
and values of dilaton coupling, 
the difference does not appear in these quantum mechanics. 
Moreover we studied the boundary states in moduli space. 
For $C \geq 3/4$, where $C$ is the coefficient of inverse square potential 
in the Hamiltonian, there is no bound state. 
For $-1/4\leq C <3/4$, that is, the cases of 
spatial dimension $N=3$ with any value of dilaton coupling $a^2$ 
and $N=5$ with $a^2>1$, 
we found the bound states and corresponding eigenfunction. 

Further studies are left following analysis. 
The obtained bound states are restricted by spatial dimensions, 
values of dilaton coupling and angular momentum. 
In the case of arbitrary dimension, dilaton coupling and angular momentum, 
the bound states must be considered. 
Also the moduli space structure is different according to the spatial dimensions 
and values of dilaton coupling. 
In the work of \cite{KSKS}, we discussed that these difference of moduli space 
structure affect quantum scattering process. 
Then we considered the quantum effect of the moduli space structure 
in the WKB approximation. 
We would like to discuss the scattering process 
with quantum states obtained on this work, more precisely.

\end{document}